\documentclass{aa}  
\usepackage[T1]{fontenc}
\usepackage[english]{babel}
\usepackage{natbib}
\usepackage{ulem,xcolor}
\usepackage{amsmath, graphicx}
\usepackage{soul}
\usepackage{nameref}
\usepackage{float}

\newcommand{\degree}{\textsuperscript{o}}

\begin{document}

\title{Three-dimensional mapping of coronal magnetic field and plasma parameters \\ in a solar flare}
\titlerunning{3D reconstruction  2021-05-07}

  \author{Tatyana Kaltman\inst{1}
          \and
         Sijie Yu\inst{2}
          \and
            Gregory D. Fleishman\inst{1,2}
           \and
          Daniel F.\ Ryan\inst{3}  
          }
\institute{Institut f\"ur Sonnenphysik (KIS), Georges-Köhler-Allee 401 A, D-79110 Freiburg, Germany
\and
Center For Solar-Terrestrial Research, New Jersey Institute of Technology, Newark, NJ 07102, USA
\and
University College London, Mullard Space Science Laboratory, Holmbury St Mary, Dorking, Surrey RH5 6NT UK}

 \date{}

\abstract
{}
{Diagnosing solar flare conditions is essential for understanding coronal energy release. Using combined microwave and X-ray data, we aim to reconstruct 3D maps of the magnetic fields and plasma parameters in the SOL2021-05-07 flare.}
{We used imaging spectroscopy from the Expanded Owens Valley Solar Array (EOVSA) to derive spatial maps of the magnetic field strength, as well as the thermal and nonthermal electron densities, along with the power-law index of nonthermal electrons via gyrosynchrotron modeling. Simultaneous X-ray observations from   Hinode/X-Ray Telescope (Hinode/XRT) and Solar Orbiter/Spectrometer Telescope for Imaging X-rays (SolO/STIX), taken from different vantage points, enable a stereoscopic reconstruction of the flaring loop. By correlating the positions of microwave and thermal X-ray sources, we associated the 3D coordinates with the microwave-derived plasma parameters.}
{We derived observational 3D maps of magnetic field strength, Alfvén speed, and plasma beta in a flaring volume, revealing a magnetically dominated environment. These spatially resolved diagnostics provide valuable constraints for models of magnetic reconnection and flare dynamics, representing a step toward a realistic 3D characterization of energy release in solar eruptive events.}
{}


\maketitle

\textit{Published in Astronomy and Astrophysics, 707, A158 (2026)}\\
\textit{\url{https://doi.org/10.1051/0004-6361/202557507}}

\section{Introduction}
Solar flares and many other solar transient phenomena are driven by free magnetic energy, which is a dominant form of energy in the corona \citep{Priest2002A&ARv..10..313P,Shibata2011}. Therefore, our understanding of the three-dimensional (3D) distribution of the magnetic field in the  corona during  flares is crucial for unraveling the physics of these transient phenomena. 

The standard solar flare model  \citep[CSHKP,][]{Carmichael1964,Sturrock1966,Hirayama1974,Kopp1976} suggests that a reconnecting X-point or implied current sheet are the key sites of the magnetic reconnection and particle acceleration; whereas recent observations \citep{Krucker2010,Fl_etal_2011, Chen2020,2022Natur.606..674F} have reported a looptop or cusp region to be the main energy release  or particle acceleration sites. Therefore, devising a 3D picture of the magnetic field and plasma is needed to localize the reconnection region, track energy release and transport, and link accelerated particles to their sources. It could also reveal the presence of current sheets, plasmoids, and flaring loops. This 3D picture is important to constrain  magnetohydrodynamic (MHD) simulations \citep{Janvier2015SoPh..290.3425,Toriumi2019LRSP...16....3T}, improving our understanding of reconnection dynamics and flare evolution \citep{Jiang2016,Arnold2021, Guo2024,2023ApJ...946...46I,2023ApJ...952..136A}.

However,  measurements of the coronal magnetic field in flares remain challenging and rare. The coronal magnetic field is often quantified using indirect approaches such as nonlinear force-free field (NLFFF) extrapolations. These models extrapolate the magnetic field from photospheric magnetograms under the assumption that the corona is approximately force-free. Although it is widely used, NLFFF extrapolations have certain limitations \citep{2006SoPh..235..161S,2012LRSP....9....5W,2009ApJ...696.1780D} and they can be entirely erroneous during episodes of major restructuring in the magnetic field that accompany large solar flares \citep{2020Sci...367..278F,2023ApJ...952..136A}. 

A new and so far unique way of inferring evolving magnetic field directly in the flare volume  is based on the microwave imaging spectroscopy of nonthermal emission produced by energetic electrons accelerated in the solar flare that produce gyrosynchrotron emission, while moving in the ambient magnetic field \citep{Gary_etal_2013}.
Broadband imaging instruments such as  EOVSA   \citep[]{Gary2018} provide spectral imaging across a broad frequency range (1–18 GHz),  enabling spatially resolved measurements of magnetic field over the flaring volume \citep{2020Sci...367..278F,2026NatAs.tmp...10F}.
\begin{figure}[hbtp!]
    \centering
    \includegraphics[width=0.98\linewidth]{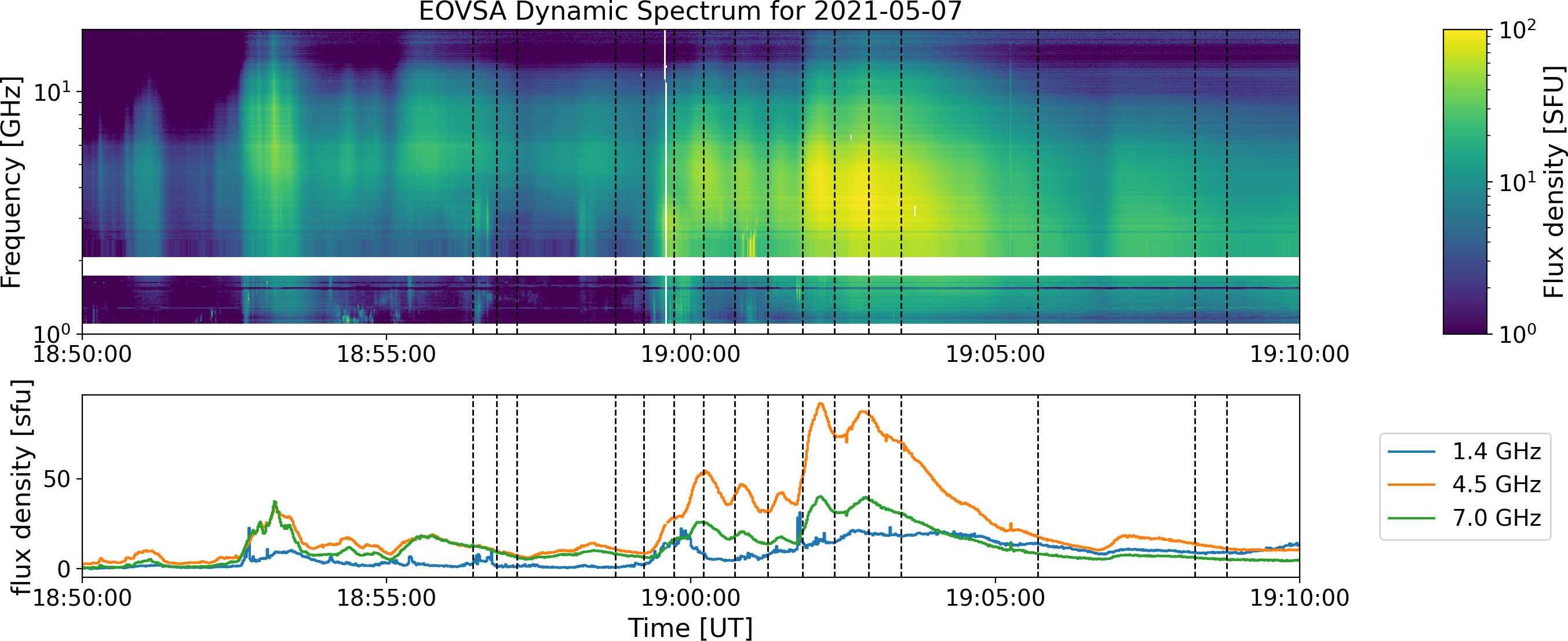}
    \caption{ Top: Background-subtracted dynamic total power radio spectrum of the 2021-05-07 solar flare observed by EOVSA (top), color scale shows the flux density in  solar flux unit (sfu).  Bottom: Time profiles at selected frequencies illustrate the burst evolution.   The black dashed lines mark the overplotted times of XRT observations. 
\label{fig:Dynamic_spectrum_EOVSA_HXR_lines}}
\end{figure}

Despite their advantages, microwave observations provide only a two-dimensional (2D) map of brightness temperature, offering no direct information about the source’s position along the line of sight (LOS). This limitation complicates efforts to reconstruct the full  3D volume of the flare, necessitating complementary observational techniques or model assumptions.

To address this, stereoscopic X-ray observations from multi-viewpoint missions can be used to determine the 3D positions of X-ray sources \citep{ryan2024a}. When the same emission region is detected in both microwave and X-ray data, the X-ray-derived geometry can serve as a proxy for 3D localization of microwave-emitting regions. This provides a 3D distribution of the magnetic field along with the plasma in flaring regions.

\cite{ryan2024a} presented the first 3D reconstruction of a thermal X-ray loop-top source during the Geostationary Operational Environmental Satellite (GOES) M3.9-class 2021-May-07 flare, using coordinated STIX and Hinode/XRT observations. Their reconstruction yielded refined  time-dependent estimates of the source’s volume  (2-5)$\times10^{27}$\,cm$^3$, plasma density  (4-7)$\times10^{10}$\,cm$^{-3}$, source mass  (2-5)$\times10^{14}$\,g, and thermal energy  (0.75-1.5)$\times10^{30}$\,erg, while demonstrating that conventional 2D approaches may overestimate the actual source size. 

In this study, we analyze the 2021-05-07 solar flare \citep[see, e.g.,][]{ryan2024a,2024ApJ_Mondal} using EOVSA observations to recover the magnetic field values in the   soft X-ray (SXR) source studied stereoscopically by \cite{ryan2024a}.
The 3D geometry derived from stereoscopic X-ray data is then used to reconstruct the spatial structure of the magnetic field, Alfvén velocity, and plasma beta in the flaring region.

\section{Observations} 
\label{sec:3d}

\subsection{Solar Orbiter/STIX and Hinode/XRT }
To estimate the 3D locations of the radio emission, we employed the results of \cite{ryan2024a}, who analyzed this same event.
They performed a 3D volumetric reconstruction of the X-ray emitting plasma as a function of time at 16 time frames 
during the flare.
This was achieved by leveraging Solar Orbiter's \citep{solo} -97\degree\ separation from the Sun-Earth line to combine cotemporal images from Solar Orbiter/STIX \citep{stix} and the Hinode/XRT Be-thick filter \citep[][]{hinode,xrt} via an elliptical tie-pointing technique \citep{Inhester2006,Horwitz2002}. 

\subsection{EOVSA}
\label{S_EOVSA}
EOVSA \citep{Gary2018} provides imaging spectroscopy at 1s cadence across 451 frequencies from 1 to 18 GHz. Raw data were retrieved from the public archive and calibrated using standard procedures, including delay, bandpass, and gain corrections. The self-calibration \citep{1999Cornwell}   was applied using the burst peak and extended to the full observational interval 18:52 -- 19:10 UT.

Imaging was performed with a 4-second cadence across 48 frequency bands ranging from 1.25 to 17.16 GHz. During the imaging reconstruction (using CLEAN), a circular restoring beam (defining the angular resolution of the image) was applied, with the full width at half maximum (FWHM) set to 60$'' \nu^{-1}$ for frequencies below 12 GHz and fixed at 5$''$ for frequencies above 12 GHz, where $\nu$ represents the observing frequency in GHz. Total power calibration was performed by scaling the image-integrated fluxes to match single-dish measurements for each frequency band. For each map, we computed the  root-mean-square (rms) value, which is typically used as a measure of error for all pixels of the given map. This approach overall works well, although it might overestimate errors in some locations with weak, but reliably measured brightness.

The temporal evolution of the microwave burst is illustrated in Fig.~\ref{fig:Dynamic_spectrum_EOVSA_HXR_lines}, which shows the background-subtracted dynamic spectrum (top) and light curves at selected frequencies (bottom). Dashed vertical lines mark the times of XRT observations used for stereoscopic reconstruction.

\begin{figure*}[htbp!]
    \centering
     \includegraphics[width=0.8\linewidth]{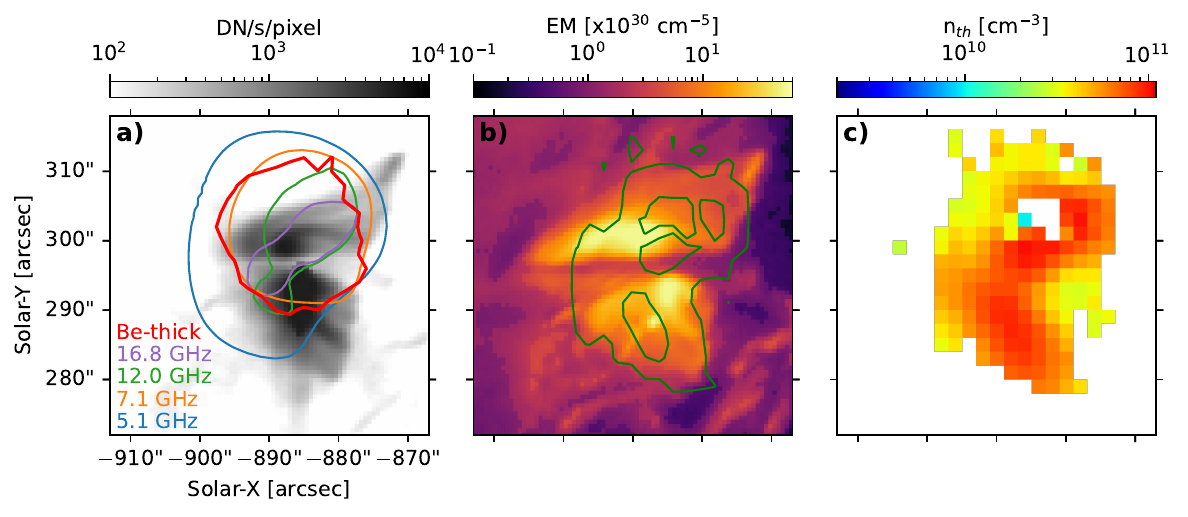}
    \caption{
     Panel a: AIA 94 Å image at 18:56:59 UT showing the flaring region. The EUV flux tube system is partly occulted by the preexisting cool filament (a horizontal absorption feature).  Thin colored contours represent 20\% of EOVSA microwave brightness at  four frequencies indicated in the legend. 
     The thick red contour shows 20\% level of the soft X-ray source observed by Hinode/XRT (Be-thick filter) at 18:56:25 UT. Panel b: Total emission measure of the flare region, with the thin green contour marks the region of enhanced thermal density at the levels of 3.6$\times10^{10}$\,cm$^{-3}$ and $8\times10^{10}$\,cm$^{-3}$, as shown in the bottom panel of Fig.~\ref{Fig:B_n_med}. 
    Given the peak $EM$ value of $\sim2\times10^{31}$\,cm$^{-5}$ and a loop depth of $\sim20\arcsec$ ($\sim$14\,Mm, roughly matching the spatial scale in the plane of the sky), the inferred peak electron density is approximately $1.2\times10^{11}$\,cm$^{-3}$, consistent with the $n_{th}$ values inferred from the microwave spectral fitting and shown in panel (c); this is also the same as the one shown in the bottom panel of Fig. \ref{Fig:B_n_med}.  Note:  most of the individual EOVSA images (contours in panel a) do not have loop shapes matching EUV loops; however, the thermal number density inferred from the multifrequency spectral fitting reveals a loop-like structure (shown in panel c), matching the AIA-derived EM map in panel b.
\label{fig:aia_131_EOVSA_XRT}}
\end{figure*}

\subsection{SDO (AIA)}
 Extreme-ultraviolet (EUV) observations from the Atmospheric Imaging Assembly (AIA) aboard the Solar Dynamics Observatory (SDO) were employed to investigate the flare structure (see Fig.\,\ref{fig:aia_131_EOVSA_XRT}) and its evolution over time. The 12-second cadence dataset was calibrated and aligned using the \texttt{aia\_prep.pro} routine provided in the SolarSoftWare (SSW) distribution \citep{2012SoPh..275...17L}. A differential emission measure (DEM) analysis is carried out using the regularized inversion method \citep{2012A&A...539A.146H}, applied to imaging data from six SDO/AIA EUV bands (94\,\AA, 131\,\AA, 171\,\AA, 193\,\AA, 211\,\AA, and 335\,\AA). The analysis yields the distribution of the LOS–integrated total emission measure,
$$
\xi(x,y) = \int n_e^2(x,y)\,dz,
$$
in the plane of the sky shown in Fig.\,\ref{fig:aia_131_EOVSA_XRT}b. This emission measure matches well the thermal number density inferred from the microwave spectral fitting (Fig.\,\ref{fig:aia_131_EOVSA_XRT}c).

\subsection{Inferring the 3D locations of radio emission}
\label{S_3D_rad_loc}

\begin{figure}\centering
\includegraphics[width=0.5\linewidth]{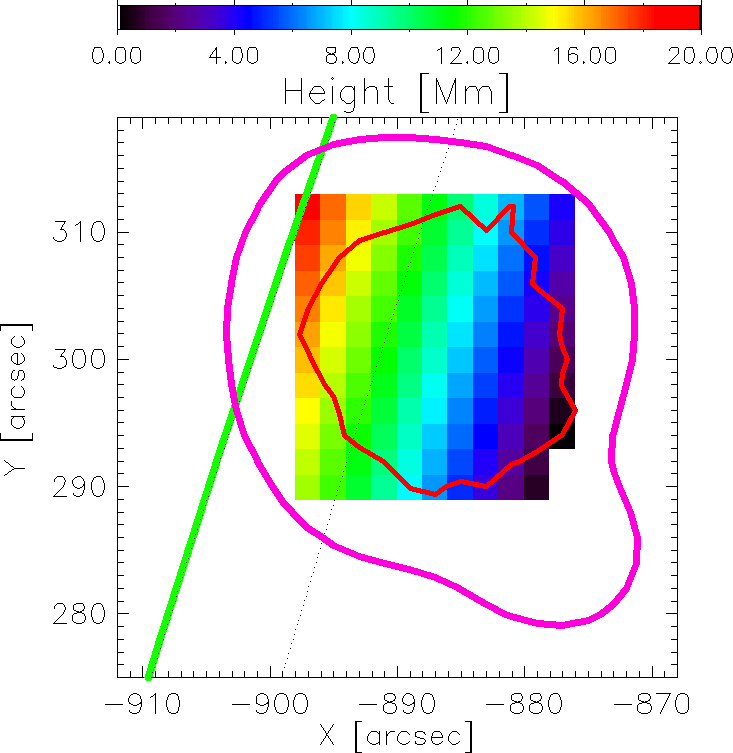}
\caption{Heights of the  SXR source  in the plane made by the solar rotation axis and the geometrical center of the SXR source obtained by \citet{ryan2024a} from 
the 3D stereoscopic reconstruction of the joint STIX/XRT data. The green line represents the solar limb. The red contour shows  Hinode/XRT 20\% brightness at 18:56:25\,UT (as in Fig.\,\ref{fig:aia_131_EOVSA_XRT}a). The violet contour shows 10\% of the 5.79\,GHz EOVSA image taken at 18:55:44\,UT.  
\label{Fig:XRT_heights}
}
\end{figure}

A comparison of the EOVSA and XRT Be-thick images (Fig.~\ref{fig:aia_131_EOVSA_XRT}) indicates that the core of the radio emission originates along the same (or very similar) lines of sight as the X-ray emission.  In addition, comparison of X-ray- and microwave-derived source parameters presented in Sect.\,\ref{S_reg_parm_map}, reveals an almost perfect match of the thermal number densities inferred from these two independent diagnostics. This implies that the microwave emission originates from the same volume as the X-ray one. However, the microwave diagnostics  do not resolve the source along  the LOS and, instead, they recover the means of the source parameters weighted with  the spatial distribution of the nonthermal electrons. Accordingly, we approximated the 3D locations (Fig.~\ref{Fig:XRT_heights}) of the radio source as lying in a plane passing through the center of the X-ray source  at the given time derived in \citet{ryan2024a} and the Sun's rotational axis.  The heights shown in Fig.~\ref{Fig:XRT_heights} were calculated from the intersections of the EOVSA lines of sight and this 2D plane. Figure~\ref{Fig:XRT_heights} displays positive heights (i.e., above the solar surface level) over a rectangular area that fully inscribes the 20\% level of the XRT X-ray brightness. 

To quantify the uncertainty in the 3D source heights, we estimated representative upper limits based on the geometry of the elliptical cross-sections used in reconstruction \citep{ryan2024a}. Specifically, for each cross-section, the uncertainty was defined as half the vertical distance between the two intersections of the ellipse boundary and the LOS from EOVSA through the ellipse center. This approach provides a conservative estimate of the maximum height variation within each cross-section. The resulting distribution of height uncertainties spans from ±0.5 Mm to ±3 Mm, with both the mean and median approaching ±2 Mm. Although some 3D positions have uncertainties below 0.5 Mm, and a few reach up to 2.9 Mm, we adopted a representative uncertainty of ±2 Mm for all reconstructed positions.

\begin{figure*}[hbtp!]
\centering
\includegraphics[width=0.98\textwidth]{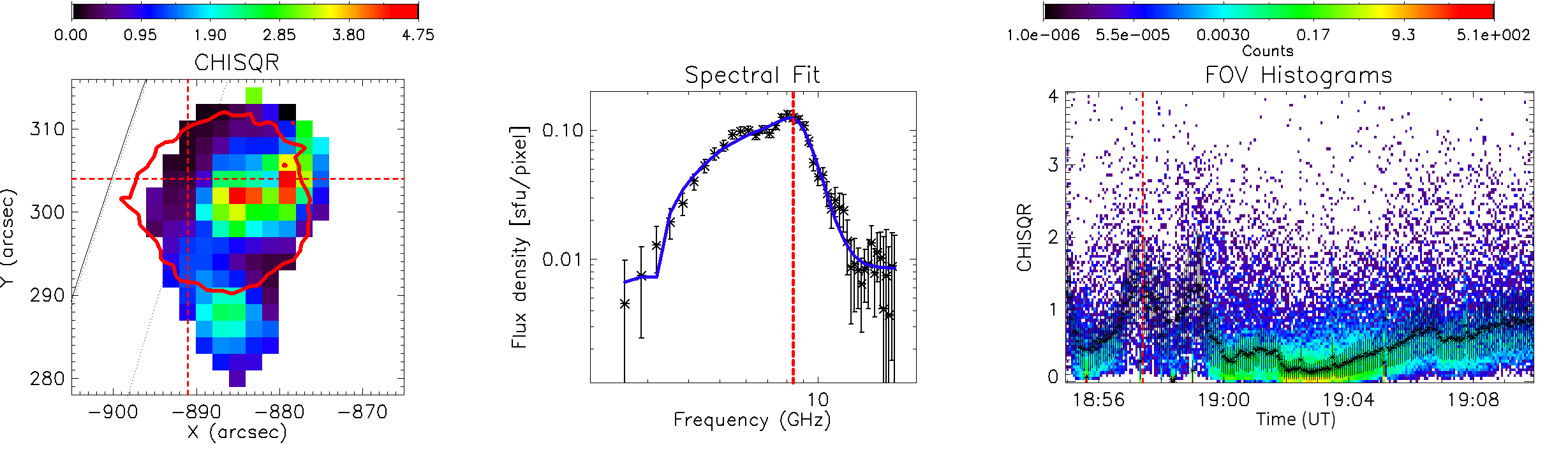}
\caption{Goodness of the spectral fit. Left: $\chi^2$  (=reduced Chi$^2$) map for the time frame of 18:57:24\,UT. Red contour shows the XRT brightness contour computed for 18:57:08\,UT. Middle: Example of the spatially resolved spectrum from the pixel shown by the red cursor (i.e., the intersection of the two red dashed lines) in the left panel and corresponding good spectral model fit shown by the blue curve.  The vertical dashed red line marks the spectral peak that demarcates the optically thick (to the left) and thin (to the right) parts of the spectrum. Right: Evolving 2D histogram of the $\chi^2$ metric (colored dots) and the corresponding median values (black symbols with error bars). This demonstrates that the spectral fits are mainly acceptable.  \label{Fig_chi-2}
}
\end{figure*}

\section{Spectral model fitting}
\label{S_Sp_fitting}

Here, we apply the method of microwave imaging spectroscopy \citep{Gary_etal_2013,2020Sci...367..278F, 2022Natur.606..674F,2026NatAs.tmp...10F} to infer the spatial distributions and temporal evolution of plasma parameters in the 2021-May-07 solar flare. Using EOVSA observations  we constructed multifrequency images of the microwave-emitting region, forming an evolving 3D data cube in frequency and space.  At each spatial pixel of the brightness maps, we fit the observed spectrum with a homogeneous gyrosynchrotron emission model  using a dedicated tool called GSFIT \citep{2020Sci...367..278F}. This uniform source represents a bar with $2\arcsec\times2\arcsec$ front area (prescribed by the pixel size) and 8\arcsec\ depth (of the order of the individual features in the flare images; see Fig.\,\ref{fig:aia_131_EOVSA_XRT}). For this spectral model fitting, we adopted an isotropic single power-law model for the nonthermal electrons with fixed $E_{\min}=10$\,keV and $E_{\max}=5$\,MeV. We also fixed the temperature of the ambient plasma at $T=20$\,MK to exclude noticeable free-free absorption as our trial spectral fit runs with free temperature showed that the fit results are not sensitive to $T$ provided it is reasonably large. Therefore, the free fit parameters are:  the nonthermal number density, $n_{nth}$, and power-law index, $\delta$, the thermal density, $n_{th}$, the magnetic field, $B$, and its angle, $\theta,$ to the LOS. Typically, $\theta$ is not well-constrained from the fits; however, we found that it is important to keep it free to permit enough flexibility for the fit to find a valid solution.  
By repeating this analysis over successive time steps, we obtained evolving parameter maps that reveal the dynamics of the flare, including the magnetic field strength, $B$, the thermal plasma density, $n_{th}$, and the parameters of the nonthermal electron distribution.

\subsection{Spectral fitting results} 
\label{S_sp_fit_res}

We performed the bulk spectral model fitting for most of the flare duration, 18:54:56--19:09:56\,UT selected to fully cover all available 3D SXR stereoscopic reconstructions, over all pixels with a significant microwave flux integrated over the spectrum (above 0.5\,sfu$\cdot$\,GHz). Figure\,\ref{Fig_chi-2} illustrates the overall goodness of the fits. Left panel: Fitted $\chi^2$  (=reduced Chi$^2$) map for a selected time frame (18:57:24\,UT) that  is centered around one and  ranges between small values (well below 1) and 4.75. This illustrates a combination of rather good and ``not-so-good" fits, which might serve an indication that the observed spectrum deviates from that from a uniform source at the corresponding locations. The cases with very small $\chi^2$ indicate that the errors are overestimated in certain locations; see Sect.\,\ref{S_EOVSA}. An example of a very good fit ($\chi^2=0.25$) that closely matches literally all spectral data points is shown in the middle panel. This match confirms that the shape of the fitted spectrum is consistent with the one from a uniform source. In other locations (shown in green and, especially, yellow and red colors), the observed spectra (not shown) have more complex shapes; namely, they have more than one spectral peak or a shallow slope(s), inconsistent with the isotropic uniform source model. In such cases, the spectral fit can randomly pick up one of the sources ignoring the others, which results in enhanced $\chi^2$ metric and ``jumps'' of the fit parameters between these two sources located on the same LOS. Right panel: Evolving 2D histogram of the $\chi^2$ distribution on top of which we overplotted the evolving median values with their corresponding uncertainties. This plot demonstrates that although there are many cases with too high $\chi^2$, we have an acceptable spectral model fit ($\chi^2\lesssim1.5$) in at least half of the valid pixels in each time frame.

\begin{figure*}[hbtp!]\centering
\includegraphics[width=0.93\textwidth]{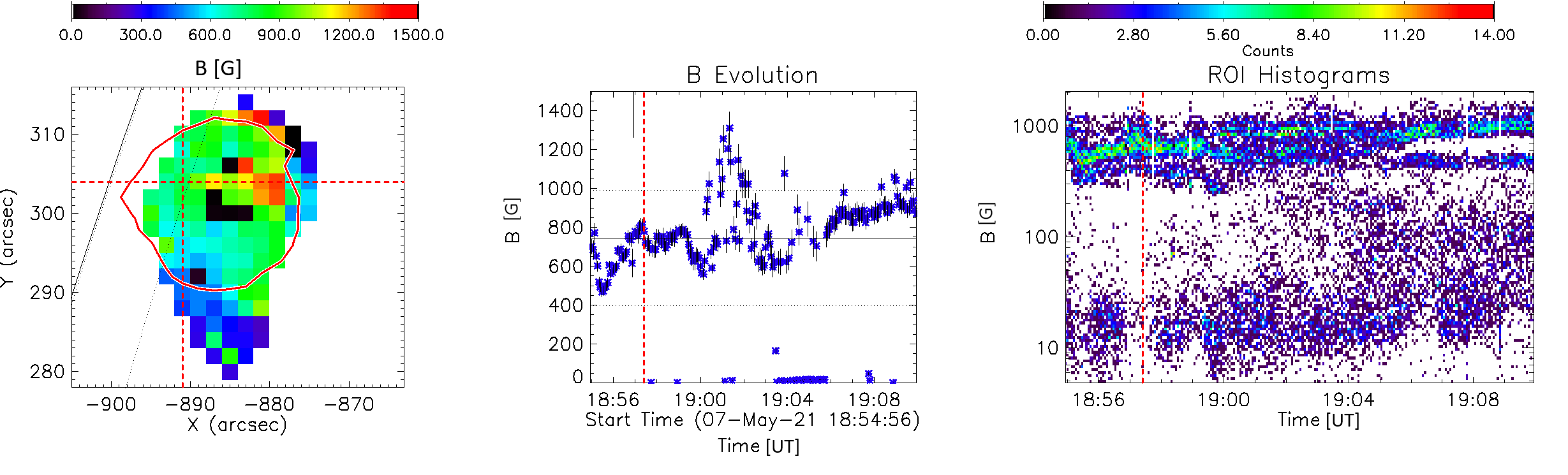} 
\includegraphics[width=0.93\textwidth]{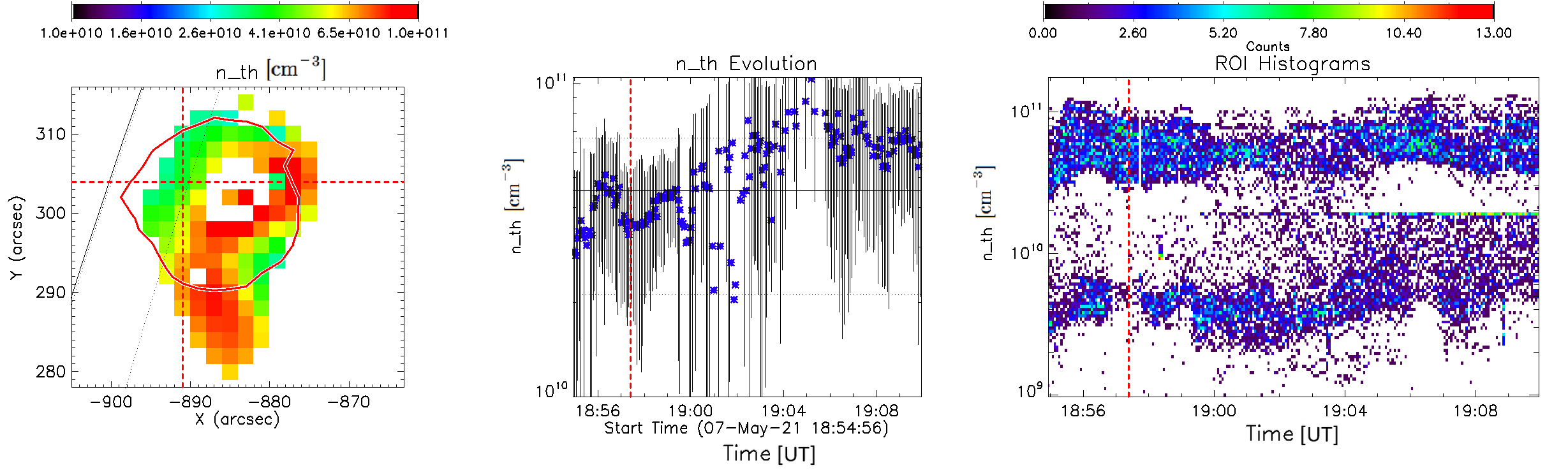} 
\includegraphics[width=0.93\textwidth]{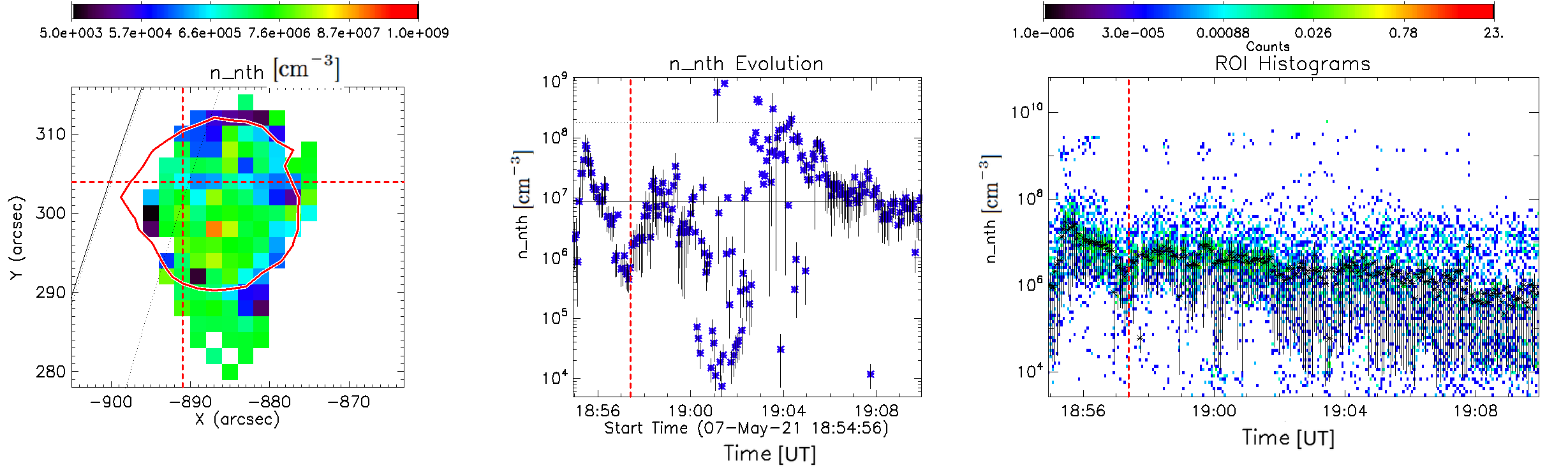} 
\includegraphics[width=0.93\textwidth]{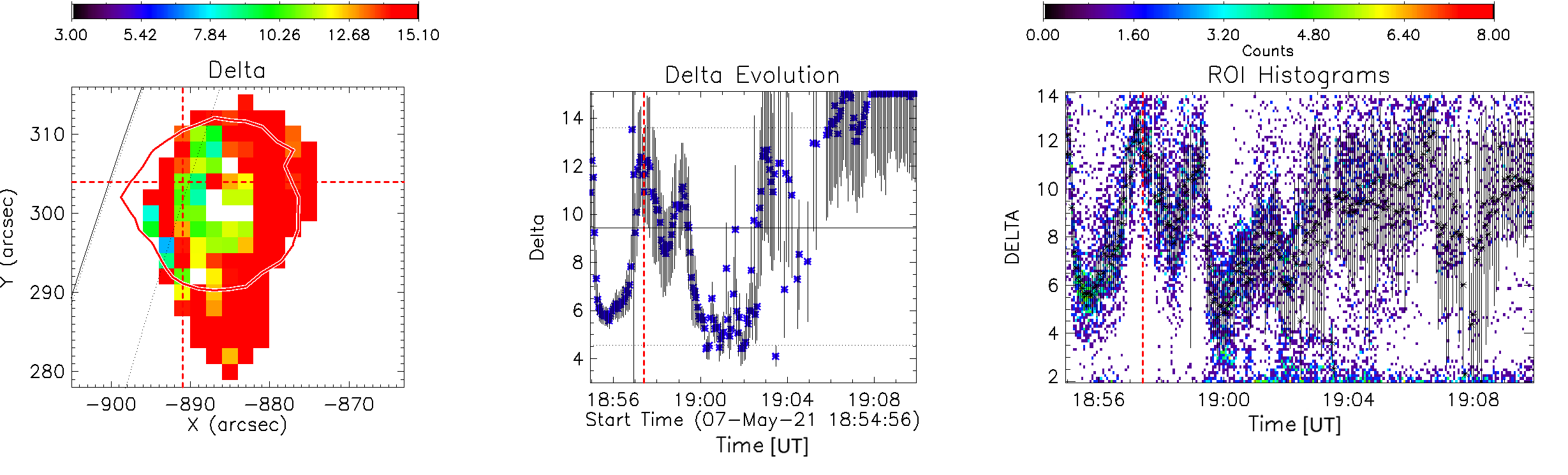} 
\caption{Parameters inferred from the spectral model fitting and their evolution. Top row: Magnetic field strength. Second row: Thermal electron number density. Third row: Nonthermal electron number density. Bottom row: Nonthermal electron power-law index. Left column: Parameter maps corresponding to the same  time frame (18:57:24\,UT) as used in Fig.\,\ref{Fig_chi-2}. Each map shows a region of interest (ROI) computed as a 20\% contour (in red) of the XRT brightness  for 18:57:08\,UT. Middle column: Evolution of the fit parameters for the cursor-selected pixel (blue symbols) along with their corresponding median values (horizontal black lines). The selected pixel shown by a cursor in the left panels demonstrates smooth spectra and good spectral fits during the entire duration of the burst. Right column: Evolution of 2D histograms of the parameter distributions within the selected ROI during the entire flare duration. For the nonthermal parameters, evolution of the corresponding median values (symbols with error bars) are shown on top of the 2D histograms to aid the eye. In both the middle and right columns, vertical dotted red lines highlight the selected time frame of the maps displayed in the left column. A horizontal stripe in 2D histogram of the thermal density at $n_{th}\approx 2\times 10^{10}$\,cm$^{-3}$ is an artifact, as explained in Fig.\,\ref{fig:Bn_hist_1D}
\label{Fig_fit_parms}
}
\end{figure*}

Figure\,\ref{Fig_fit_parms} illustrates instantaneous parameter maps and their evolution. The time frame, 18:57:24\,UT, the same as in Fig.\,\ref{Fig_chi-2}, was selected at the end of the first analyzed impulsive peak of the radio burst, when, presumably, the effect of source nonuniformity along the LOS must be minimal\footnote{If we have two (or more) sources along the LOS at a decay phase, emission from one of them  might decay slower than the other(s). Thus, at the late decay phase emission from that one  may dominate.}. All demonstrated parameter maps are rather smooth, albeit with several defects  or artifacts. The thermal number density map displays a nice loop-like structure, assumed to be tracing a hot and dense post-flare loop with  number density  within the range of $n_{th}\sim (2-8)\times10^{10}$\,cm$^{-3}$. This loop structure projects onto a bright AIA feature clearly seen in Fig.\,\ref{fig:aia_131_EOVSA_XRT}, which is partly occulted by a cool filament. The filament material is transparent for the microwave emission; thus, the loop structure is recovered using the microwave spectral model fitting.
The nonthermal number density  in this loop is  about $n_{nth}\sim 10^{6}$\,cm$^{-3}$; the spectrum of the nonthermal electrons is very soft, $\delta\gtrsim10$, while distinctly different from a purely thermal one \citep[see, e.g.,][]{Fl_etal_2015,2017NatAs...1E..85W}. Such a very tiny population of the  nonthermal electrons is expected at the phase of local minimum of the microwave brightness.

The second column in Fig.\,\ref{Fig_fit_parms} displays evolution of these parameters in the pixel shown by the cursor in the parameter maps. This pixel was selected because most of the spectra from this location during the burst duration are  consistent with a homogeneous-source model, as in Fig.\,\ref{Fig_chi-2}, permitting a meaningful spectral model fitting.  There are many other pixels, where the $\chi^2$ metrics are above 2-3, which is an indication that the uniform source model might not be sufficient to fit the data, because more than one source is likely located  along the LOS. An inspection of such cases and post-processing filtering of the fit results indicate that the uniform source fit "picks up" either one or the other source along the LOS, which is further elaborated on in Sect.\,\ref{S_reg_parm_map}. 

During the first analyzed impulsive peak, 18:54:56--18:57:24\,UT, the magnetic field in this location goes down from 700\,G to 470\,G, with the rate of $\dot{B}\sim-7$\,G\,s$^{-1}$ similar to earlier reported values \citep{2020Sci...367..278F, 2026NatAs.tmp...10F}. However, this only lasts half of minute and then slowly rises up to 810\,G with the rate $\dot{B}\sim3$\,G\,s$^{-1}$. In principle, these changes in the inferred magnetic field might represent a geometrical effect, when different portions of the LOS are illuminated by the radio-emitting nonthermal electrons in different moments, rather than the evolution "in place." However, the inferred thermal number density remains in the range of the density inferred from the X-ray data;  thus, we are likely detecting real temporal variations in $B$. Further  apparent evolution of the magnetic field represents a combination of decreases and increases overall (anti)correlated with microwave brightness peaks in time (presumably, episodes of the free magnetic energy release) superimposed on a general trend of the magnetic field increase up to $\sim$900\,G with the rate $\dot{B}\approx0.2$\,G\,s$^{-1}$. During the maximum phase of the burst, $\sim$19:00--19:04\,UT, the fit solutions for the magnetic field are not unique, even though the individual spectra are pretty smooth.

The thermal number density clusters around a few $10^{10}$\,cm$^{-3}$; the uncertainties of individual values are typically about 20-30\%. Thus, the thermal number density is well constrained by the data. The mean values show a slowly increasing trend, but the range of the variation remains within the uncertainties and, thus, they are not statistically significant. 

The nonthermal number density is clearly correlated with the microwave brightness, as expected, until about 19\,UT. Afterward, it could not be well constrained because of the aforementioned nonuniqueness of the fit at the later times. After  the episode when the fit is not unique ($\approx$ 19:00 -- 19:04 UT), the nonthermal number density gradually decreases as the burst fades out. 

The spectral index, $\delta$, demonstrates a very prominent soft-hard-soft behavior during most of the individual temporal brightness peaks. The range seen for the $\delta$ variation is between 4 and 14, which is much wider than typically reported from X-ray data \citep[e.g.,][]{Grigis2004A&A...426.1093G}.  Even though the  hard X-ray is mainly produced in the footpoints, and the microwave emission  comes from a coronal volume, which can explain different values of the indices, the difference in their ranges is striking. A possible interpretation of this dissimilarity between the X-ray and microwave diagnostics, already proposed in \cite{2025ApJ...988..260F}, is that is it likely due to their sensitivity to nonthermal electrons with different energies, which evolve differently \citep[see, e.g., ][]{2021ApJ...908L..55C}.

\begin{figure}[hbtp!]
    \centering
\includegraphics[width=0.49\linewidth]{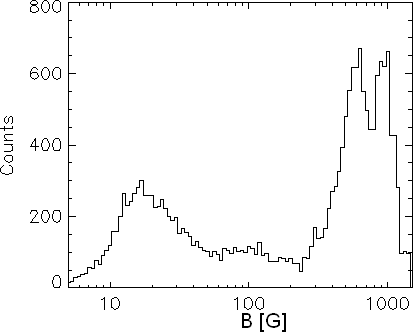}
\includegraphics[width=0.49\linewidth]{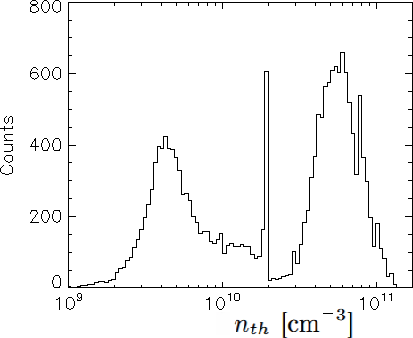}
\caption{1D histograms of the inferred magnetic field $B$ [G] and thermal number density, $n_{th}$ [cm$^{-3}$], during the entire analyzed flare duration. The strong narrow bin in the $n_{th}$ histogram at $n_{th}\approx 2\times 10^{10}$\,cm$^{-3}$ is an artifact of a restricted EOVSA spectral range, as this number density is associated with the plasma frequency of 1.25\,GHz (the smallest available frequency). 
    \label{fig:Bn_hist_1D}}
\end{figure}

These parameter trends discussed for a single pixel are consistent with general trends displayed by the third column showing evolving 2D histograms of the same parameters for all pixels inscribed within the red ROI. However, these histograms reveal additional features, which cannot be identified for an individual pixel. The histograms of the magnetic field and the thermal number density display a pronounced dichotomy (or even a trichotomy) of solutions. This is due to  aforementioned complexity of many individual spectra that show signatures of several sources along the LOS, while the uniform source spectral model can report properties of a single source only. Given that the relative contributions of these different sources to the total spectrum can vary in space and time, the spectral fitting result fluctuate between these different sources yielding  the parameter dichotomy. We note that the effect of this dichotomy is minimal in the local minimum of the microwave brightness at 18:57:00 -- 18:58:00\,UT; the middle of this interval is shown by the vertical red line. 
This dichotomy and trichotomy are further illustrated by Fig.\,\ref{fig:Bn_hist_1D}, which shows a 1D histogram of the magnetic field and thermal number density within the specified ROI during the entire analyzed flare duration.
Interestingly, the nonthermal electron population does not exhibit any dichotomy and this could possibly point to at most a single acceleration source in this event.

\subsection{ Median parameter maps}
\label{S_reg_parm_map}
In Sect.\,\ref{S_3D_rad_loc}, we note that the nonthermal microwave source projects onto the Hinode/XRT SXR source, suggesting that they originate from the same or closely related regions. However, the discovered source complexity along many LOSs implies the presence of at least two different microwave sources along those LOSs. This makes it difficult to associate any of those sources with the SXR source for which the stereoscopic 3D reconstructions are available. To address this issue, we note that (according to Fig.\,\ref{fig:Bn_hist_1D}) those two sources have notably different magnetic field and thermal number density. 
Thus, to disentangle the physical properties of those two distinct sources, we needed to create  averaged parameter maps as follows. We considered a restricted two-minute time range, 18:55:00--18:57:00\,UT, where there are many (but by far not all) good spectra, allowing for unique spectral fits. For each pixel and time frame, we sorted out the fit outcome into two cases:\ with the inferred magnetic field above or below 100\,G, based on the value that separates two main peaks in the histogram shown in Fig.\,\ref{fig:Bn_hist_1D}. Then, for each pixel with the strong magnetic field, we determined, over the corresponding time frames within our two-minute interval, the median values and standard deviations of the magnetic field and thermal number density and assigned all those values to the corresponding pixels to obtain the  median parameter maps of this ``main'' source. Then, we repeated this procedure for the pixels-over-time frames associated with the small magnetic field (in a ``secondary'' source). 

\begin{figure*}\centering
\includegraphics[width=0.48\linewidth]{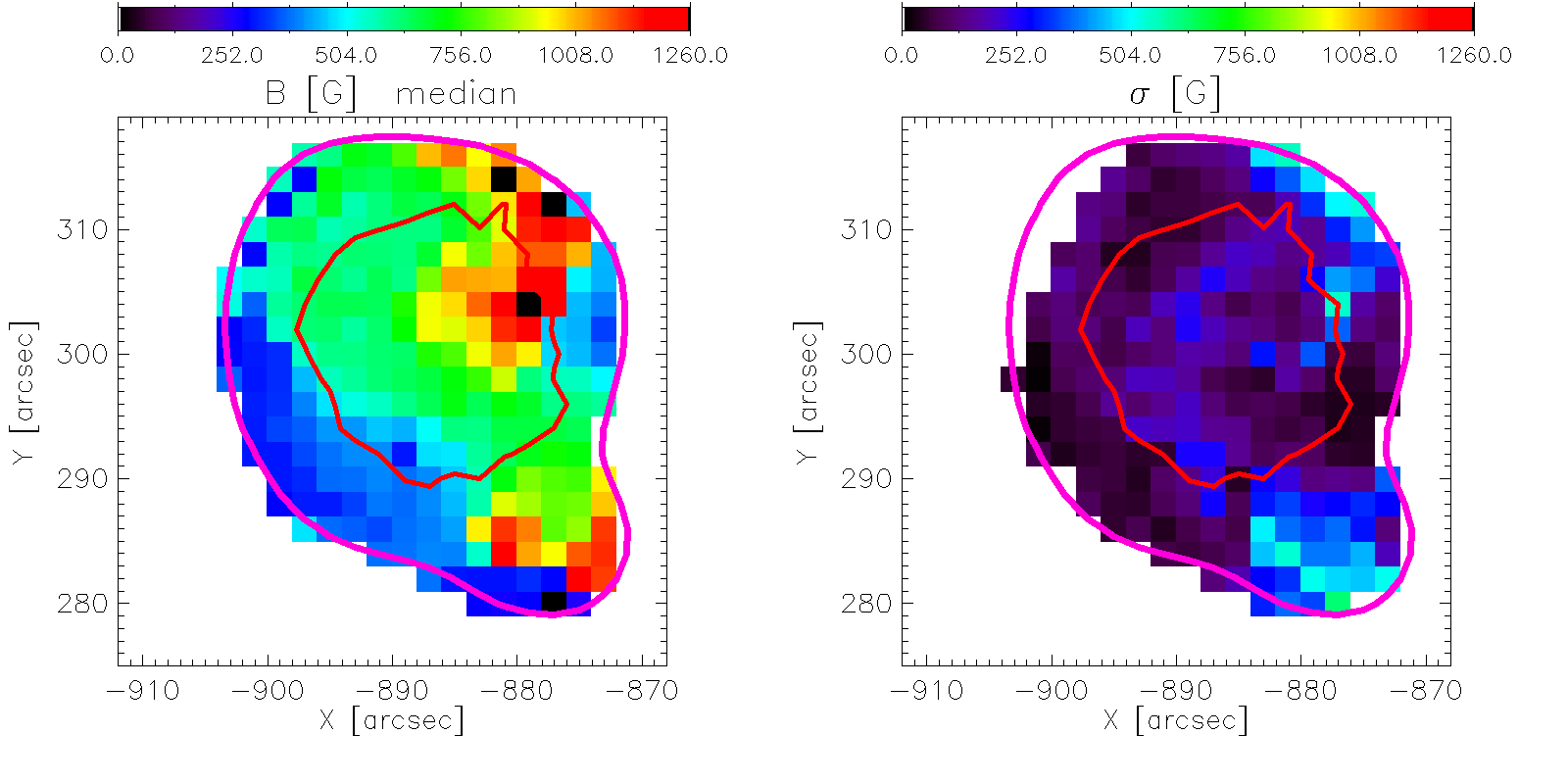}
\includegraphics[width=0.48\linewidth]{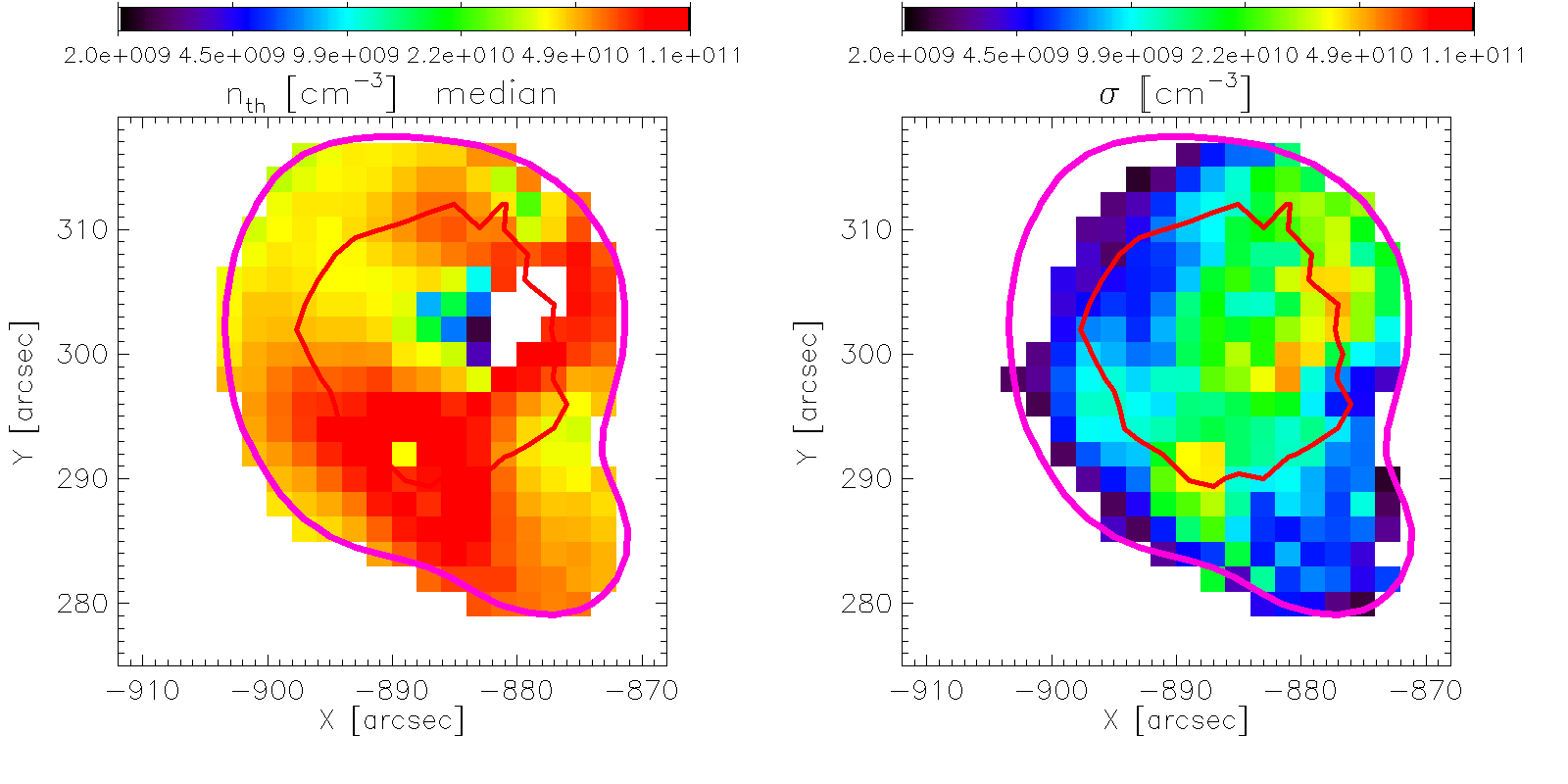}
\includegraphics[width=0.48\linewidth]{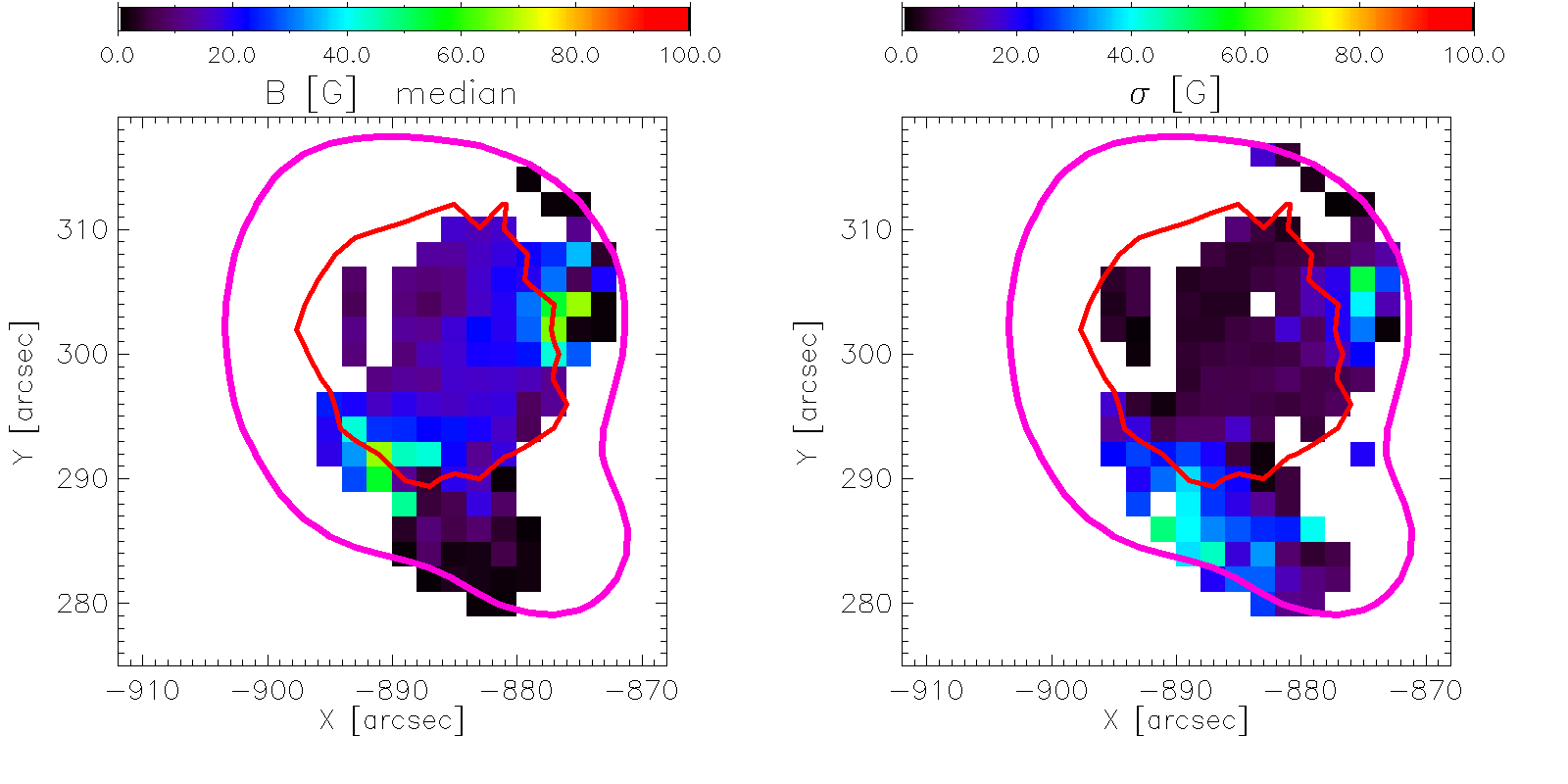}
\includegraphics[width=0.48\linewidth]{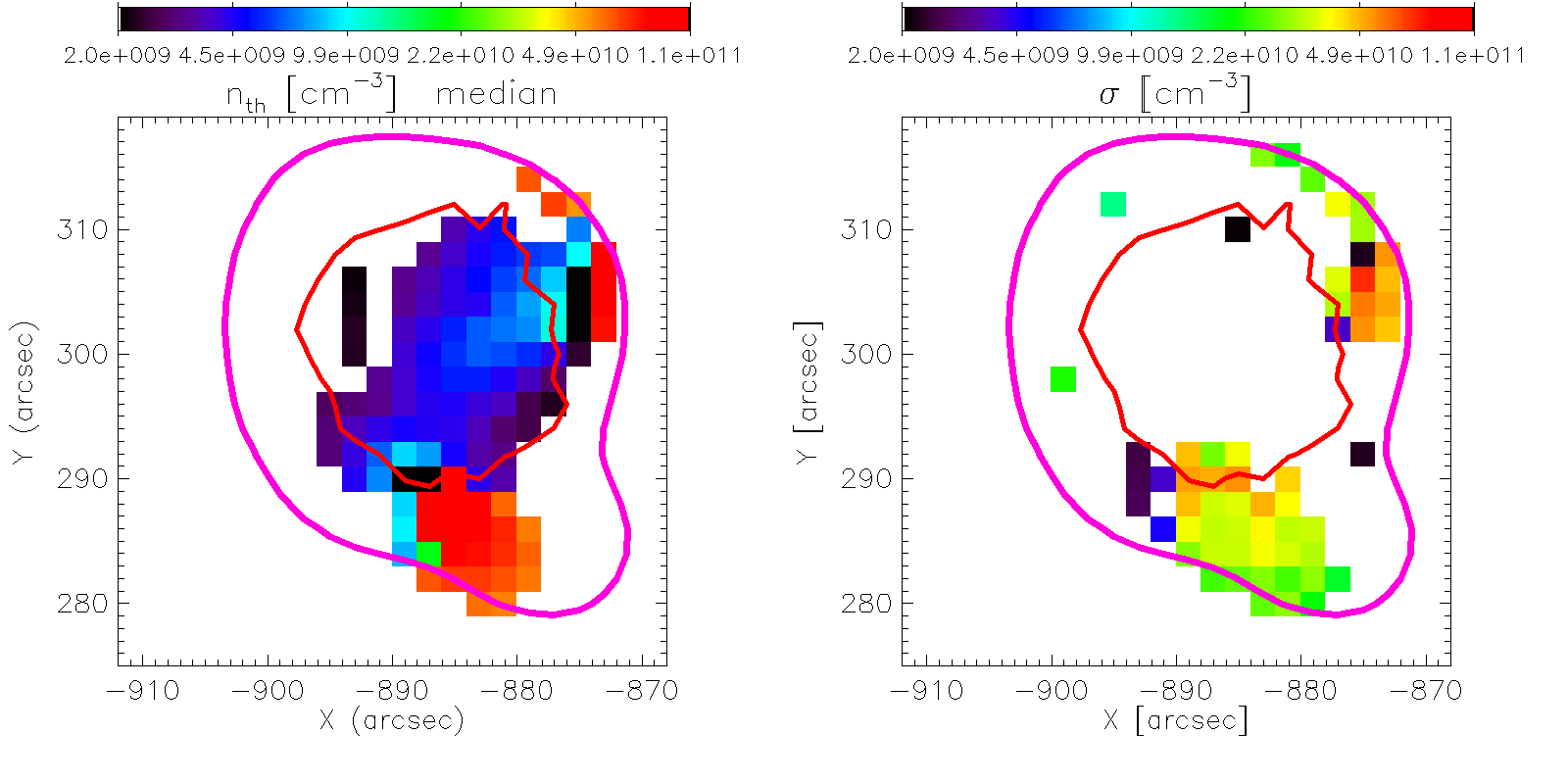}
\includegraphics[width=0.48\linewidth]{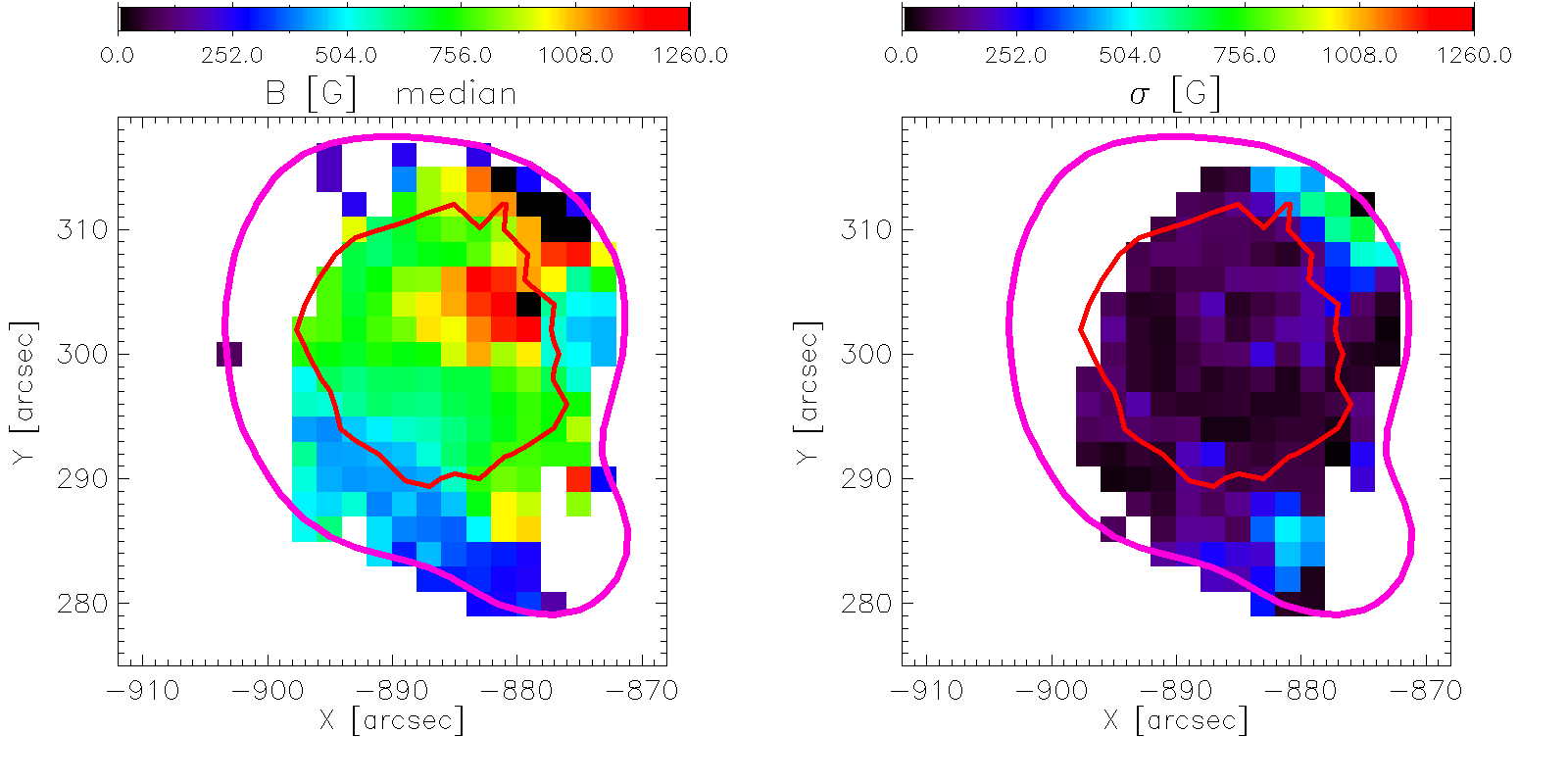}
\includegraphics[width=0.48\linewidth]{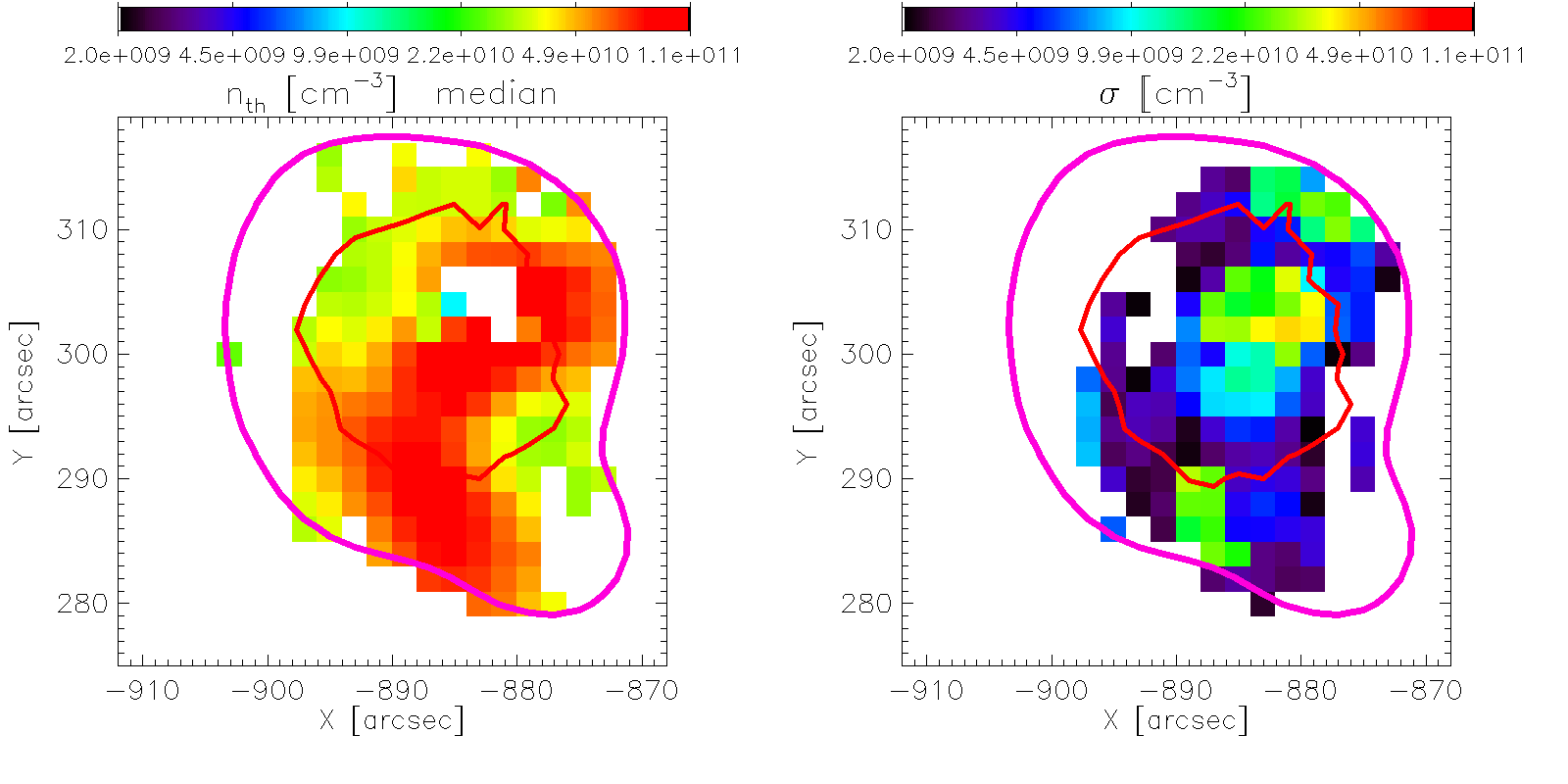}
\caption{
Maps ($44\arcsec \times 44\arcsec$) of the coronal magnetic field strength and thermal electron number density, along with their standard deviations. 
Top and middle panels: Derived over the two-minute interval 18:55:00--18:57:00\,UT for pixels with $B > 100$\,G and $B < 100$\,G, respectively. 
Bottom panel: Derived over the one-minute interval 18:57:00--18:58:00\,UT for pixels with $B > 100$\,G. The red contour shows the 20\% ROI level from Hinode/XRT at 18:56:25\,UT, and the violet contour indicates the 10\% brightness level from the 5.79\,GHz EOVSA  image at 18:55:44\,UT.
\label{Fig:B_n_med}
}
\end{figure*}

\begin{figure}\centering
\includegraphics[width=0.5\linewidth]{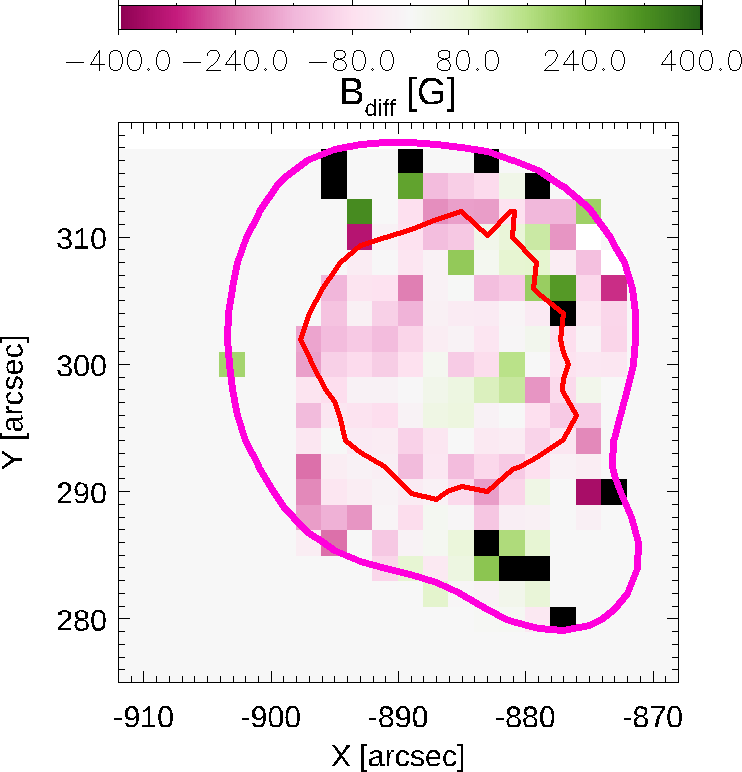}
 \caption{Map of the median magnetic field difference between two intervals: [18:55:00--18:57:00]\,UT minus [18:57:00--18:58:00]\,UT, with the median computed in both cases using only pixels where $B > 100$\,G. The red contour shows the 20\% ROI level from Hinode/XRT at 18:56:25\,UT, while the violet contour indicates the 10\% brightness level from the 5.79\,GHz EOVSA  image at 18:55:44\,UT. 
\label{Fig:map_B_differ}
}
\end{figure}

Figure\,\ref{Fig:B_n_med} shows these  median maps in the first and second rows. These maps clearly demonstrate that the strong magnetic field is associated with  overall a denser source with the same number density as observed in SXRs. The other source (shown in the second row) is  more tenuous by a factor of 10  and roughly two orders of magnitude fainter overall than the main source in the SXRs; thus, it appears to be invisible there. This permits us to associate the main source with the SXR source and assign 3D coordinates of the SXR source to the microwave inferred magnetic field values in the main source. To double check this conclusion, we applied the same approach to another one-minute interval, 18:57:00 -- 18:58:00\,UT, which is almost free from the solution dichotomy. The corresponding  median parameter maps are shown in the third row of Fig.\,\ref{Fig:B_n_med}. It is apparent that they are similar to the parameter maps of the main source obtained for the previous 2-minute interval,  validating our approach. We note that the regularized maps from the bottom row of Fig.\,\ref{Fig:B_n_med} are very similar to the instantaneous maps shown in Fig.\,\ref{Fig_fit_parms}, indicating the robust performance of the spectral model fitting during the analyzed one-minute interval. 

 The second and fourth columns of Fig.\,\ref{Fig:B_n_med} show maps of the standard deviations of the corresponding parameters, which demonstrate the statistical significance of the inferred parameters over most of the area. However, in a region where the magnetic field is strong (red colors in the upper part of the maps in the first and third rows), the thermal number density is not well constrained (see white-blue colors in the parameter maps vs. yellow-green colors in the standard deviation maps). The reason  for this behavior is well understood. The thermal number density affects the gyrosynchrotron spectrum via the dispersion relations of electromagnetic waves \citep[the so-called Razin effect; see, e.g., Sect. 9.4.2 in][]{FT_2013}. The Razin effect primarily affects the gyrosynchrotron spectrum at frequencies of $f\lesssim20n_{th}/B$. Thus, in sources with a strong magnetic field ($B\sim1200$\,G), the gyrosynchrotron spectrum is only weakly sensitive to the thermal number density at frequencies of $f\gtrsim1.5$\,GHz provided that $n_{th}\lesssim10^{11}$\,cm$^{-3}$. Therefore, the white-blue "gaps" in the thermal density maps do not imply low-density regions; rather, they point to the regions where this density cannot be constrained using the microwave data.

Figure\,\ref{Fig:map_B_differ} shows an overall modest increase (typically, by less than 100\,G with a median value 45\,G) of the magnetic field across most of the ROI between the intervals 18:55:00–18:57:00 UT and 18:57:00–18:58:00 UT. This indicates a magnetic field increase rate of about $\dot{B}\sim0.5$\,G\,s$^{-1}$. 

\begin{figure*}[hbtp!]\centering
\includegraphics[width=0.9\linewidth]{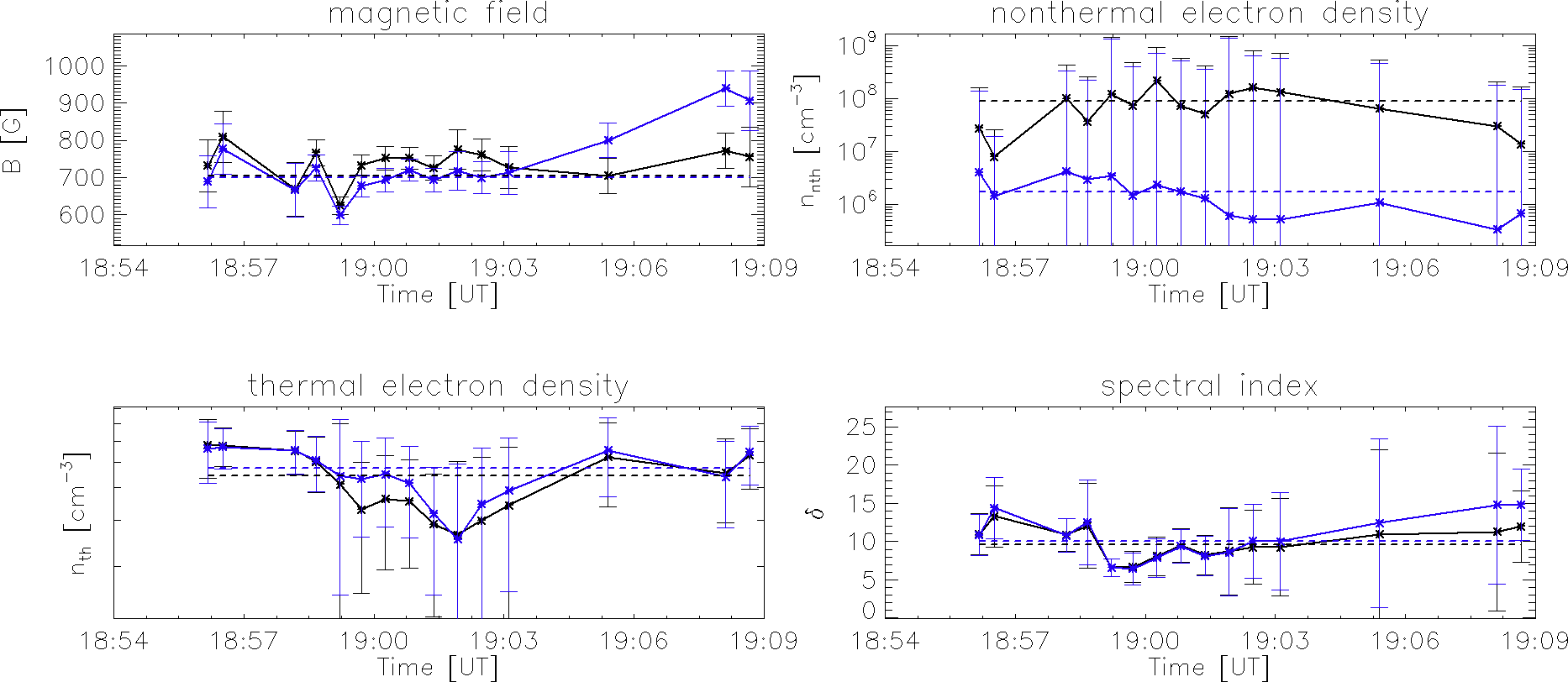}
\caption{
Temporal evolution of the mean (black) and median (blue) values of four physical parameters: magnetic field strength ($B$), nonthermal electron density ($n_{\mathrm{nth}}$), thermal electron density ($n_{\mathrm{th}}$), and electron spectral index ($\delta$), derived from radio diagnostics. The parameters are averaged over each of the 16 ROIs \citep[corresponding to 16 time frames analyzed by][]{ryan2024a}. 
Error bars denote the standard deviation across the ROIs at each time step. Horizontal dashed lines indicate the  mean and median of each parameter over the full interval. All parameters exhibit only minor variations with time, remaining within the respective uncertainties. 
\label{Fig:param_trends}
}
\end{figure*}

Below,  we capitalized on the fact that the thermal number density in a denser source inferred from the microwave data spans the range of $n_{\mathrm{th}} \sim (1{-}10)\times10^{10}$\,cm$^{-3}$, with a median value near $5\times10^{10}$\,cm$^{-3}$, which matches  the density derived from the stereoscopic SXR data analysis perfectly \citep{ryan2024a}. In that case, the density varies between approximately $(5{-}7)\times10^{10}$\,cm$^{-3}$ (their Fig.~6e); thus, we can focus on that denser source. 
This match  is consistent with the assumption that the volume filling factor is close to unity in the flaring source.  Given the absence of comparably dense plasma at other locations than the SXR source, we can conclude that the microwave and SXR emissions arise from the same physical volume. This provides a robust basis for assigning the 3D X-ray–derived coordinates to the plasma and magnetic field parameters inferred for our main source from the microwave analysis.

\subsection{Evolution of the fit parameters moments}

For each of the 16 ROIs, employed in \citet{ryan2024a} for the SXR stereoscopic reconstruction, we analyzed the physical parameters inferred from the microwave spectral model fitting described in Sect.\,\ref{S_sp_fit_res}  within a temporal window of $\pm12$~s around the corresponding time frame.  Only fit solutions with magnetic field strengths above $100~\mathrm{G}$ were included in the analysis. 

Within each ROI, over each 24-second interval, the mean, median, and standard deviation of the parameters were calculated to provide representative values and associated uncertainties. The resulting time dependences are presented in Fig.~\ref{Fig:param_trends}.

The temporal profiles of the inferred parameters show no significant systematic evolution. The median magnetic field strength ($B$) displays a slight upward trend consistent with earlier revealed increase of $\dot{B}\approx0.2$\,G\,s$^{-1}$. 
The nonthermal electron density ($n_{\mathrm{nth}}$) varies over nearly two orders of magnitude; however, the large disparity between the mean and median, together with the broad error bars, suggest the presence of outliers, rather than a global trend. The thermal electron density ($n_{\mathrm{th}}$) remains nearly constant throughout the analyzed period, aside from a short-lived dip around 19:02~UT. The spectral index ($\delta$) fluctuates within the range $\sim$(7 -- 15), with increasing uncertainties toward the end of the interval.

The weak temporal variability and close agreement between the mean and median trends of the magnetic field indicate that it did not evolve noticeably within the selected XRT ROIs. This justifies the use of short (1--2~min) temporal segments for further analysis and allowed us to average parameter values over such intervals, which provides smoother parameter maps than for any individual time frame.

These median values complemented by the source volume defined from the SXR stereoscopy, $V=5\times10^{27}$\,cm$^3$,  enable us to compute and compare the available energy and pressure of the magnetic field and thermal plasma. Specifically, for 
$B_{med}\approx750$\,G, we obtained $W_B\approx10^{32}$\,erg, while for $n_{nth, med} \approx 10^{8}$\,cm$^{-3}$, $\delta_{med}\approx8$, and adopted $E_{\min}=10$\,keV we obtain 
$W_{nth} \approx2.4\times10^{29}$\,erg, which is less than the magnetic energy by more than two orders of magnitude; thus, the inferred magnetic field is sufficient to power the flare \citep[cf.][]{2020Sci...367..278F}. 

These data also permit us to check if the derived flare parameters are consistent with the plasma confinement by the magnetic field. At the looptop edge, we have $B=306$\,G; thus, the magnetic pressure is $p_B=1120$\,erg\,cm$^{-3}$, which is a factor of 5 higher than the thermal pressure $p_{th}\approx230$\,erg\,cm$^{-3}$ obtained for the microwave derived thermal number density and X-ray derived temperature of 16\,MK \citep[][the microwave spectra are not sensitive to  such high temperatures as was noted in the beginning of Sect.\,\ref{S_Sp_fitting}]{ryan2024a}. This dominance of the magnetic pressure is sufficient for plasma confinement.

\begin{figure}[hbtp!]
\centering
\includegraphics[width=0.98\linewidth]{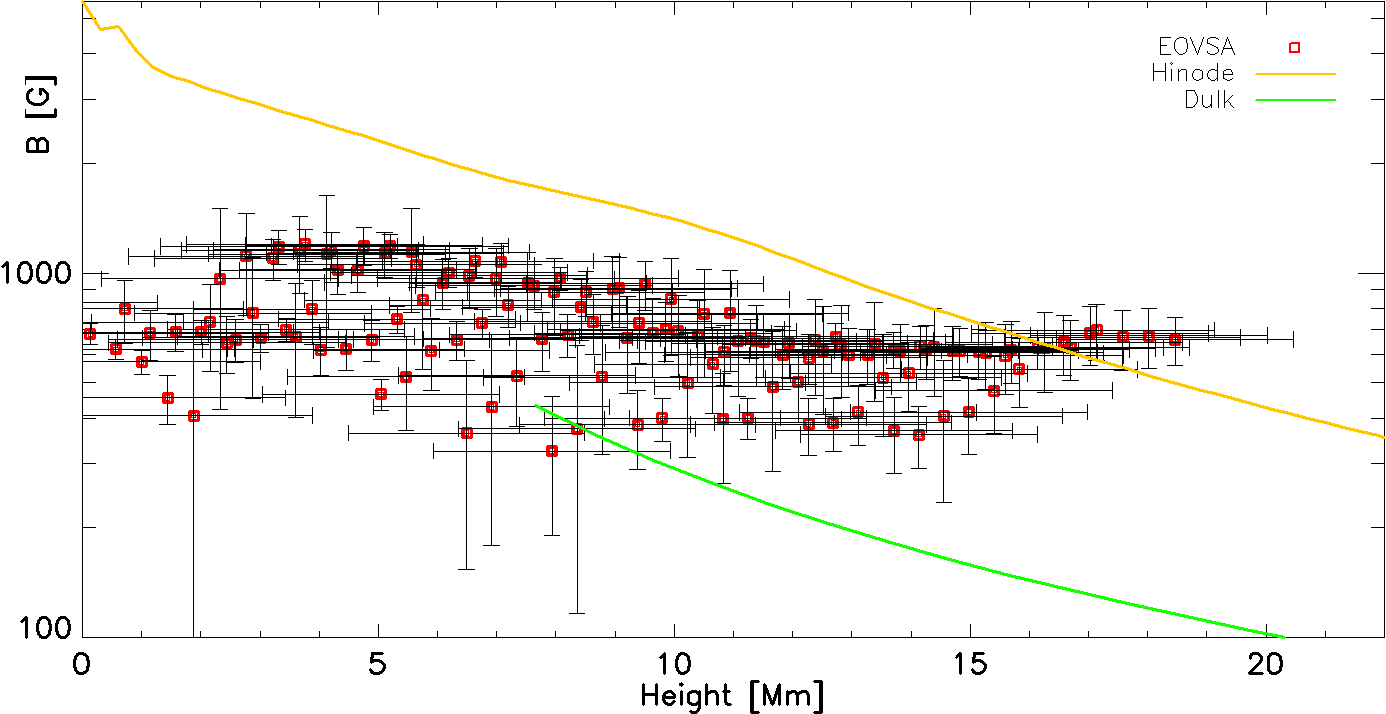}
\caption{ Scatter plot of the representative heights of the EOVSA microwave source (estimated by intersecting the EOVSA lines of sight with the plane defined by the Sun’s rotation axis and the geometric center of the SXR source; see Fig.~\ref{Fig:XRT_heights}) versus the corresponding mean magnetic field strengths. Vertical and horizontal bars indicate uncertainties in height and magnetic field strength, respectively. The green and orange lines show empirical magnetic field height profiles from \citet{2023ApJ...943..160F} overplotted for reference. The correspondence indicates that our reconstructed magnetic field values are physically plausible and lie within the expected range for active region loops.
\label{scatter_dist_B_add_Hin_Dulk}
}
\end{figure}

\subsection{Height dependence of the magnetic field}
3D coordinates of our reconstruction permit us to determine how magnetic field strength varies with height in the corona, which is important for many practical applications.
Figure~\ref{scatter_dist_B_add_Hin_Dulk} shows this dependence in the form of a scatter plot of the  median magnetic field strength versus height of the source, for the pixels within the ROI obtained from Hinode/XRT 20\% image at 18:56:24.9\,UT. This plot reveals an overall decrease of the field strength with height as expected. 

To place our results in the context of previous observations and models, we compare them with the published profiles of the coronal magnetic field dependence on height. The green line shows an empirical dependence for a ``typical'' magnetic field in active regions (AR) inferred by \citet{1978SoPh...57..279D} from a combination of different indirect techniques. Most of our inferred data are located above this line. The yellow line shows a maximum magnetic field $B_{\max}(h)$ obtained from  NLFFF extrapolation devised for AR\,12673 at 2017-Sep-06, when this AR displayed high photospheric \citep{2018RNAAS...2a...8W} and record-breaking coronal magnetic field \citep{2019ApJ...880L..29A}. By construction, the yellow line represents a (soft) upper bound of the AR magnetic field in the steady state, confirmed by a compilation of various coronal magnetic field measurements performed by \citet{2023ApJ...943..160F}.\footnote{An interactive version of their $B(h)$ plot can be found at \url{https://www.leibniz-kis.de/en/research/solar-magnetism/coronal-magnetic-field/}.} Our inferred magnetic field data are mostly below this line.

\section{3D Maps of the Alfvén speed and plasma beta}

The reconstructed 3D maps of magnetic field strength and plasma number density allow us to derive key physical parameters that govern plasma dynamics in the solar corona; most notably the Alfvén speed $v_{\mathrm{A}} = B / \sqrt{4\pi \rho}$ and the plasma beta  $\beta = 8\pi p / B^2$ \citep[see, e.g.,][]{Gary2001SoPh,Arnold2021}, where \( B \) is the magnetic field strength, \( \rho = 1.2\, m_p\, (n_{\text{th}} + n_{\text{nth}}) \) is the mass density, and \( p = 3 n_{\text{th}} k_B T \) is the thermal pressure of the plasma. Here, $m_p$ is the proton mass, while the factor \( 1.2 \) accounts for the presence of helium and heavier ions according to solar coronal abundances.
The thermal pressure, \( p \), is calculated from the inferred thermal electron number density using the ideal gas equation of state, where \( k_B \) is the Boltzmann constant and \( T = 16\,\text{MK} \) is the plasma temperature inferred from the SXR data \citep{ryan2024a}.

\begin{figure*}[hbtp!]\centering
\includegraphics[width=0.8\linewidth]{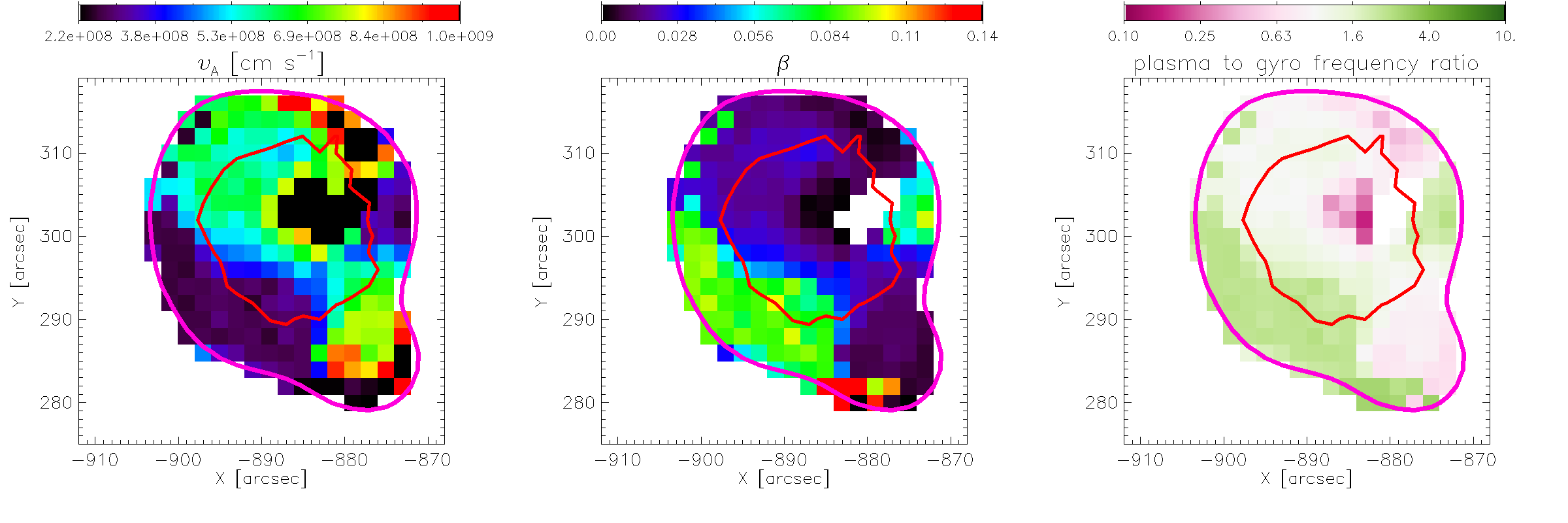}
\includegraphics[width=0.8\linewidth]{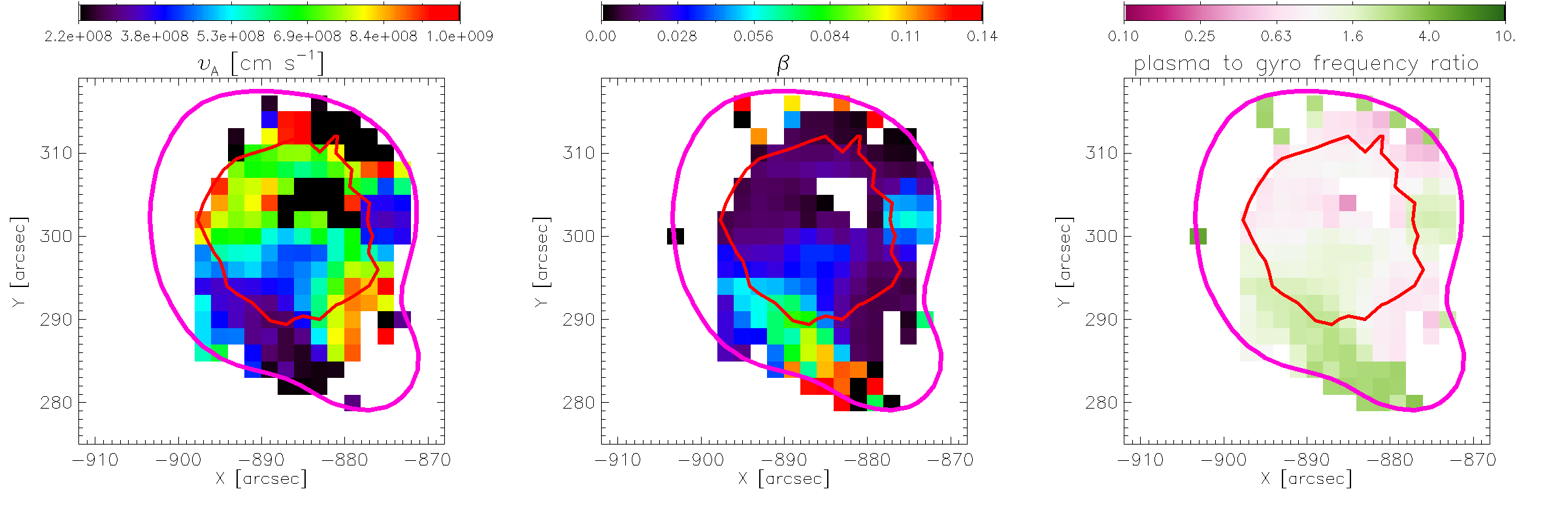}
\includegraphics[width=0.8\linewidth]{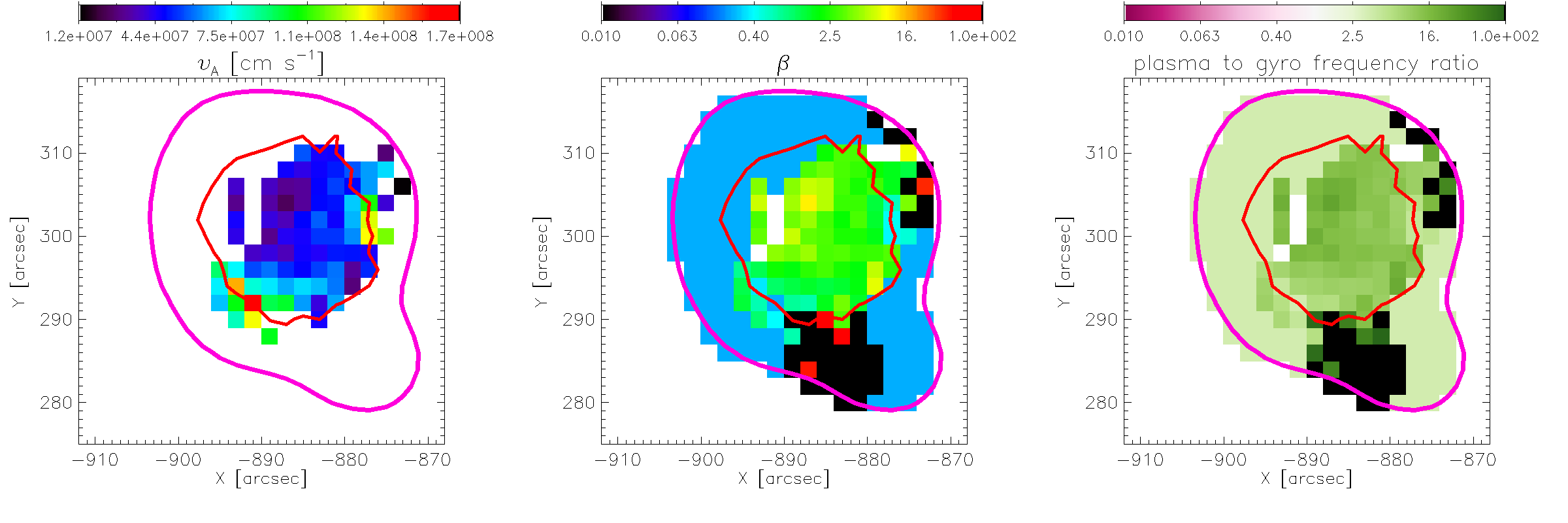}
\caption{Maps of the Alfvén velocity (left), plasma beta (center), and plasma-to-gyro frequency ratio (right) for the high-B source (top two rows, time intervals 18:55:00--18:57:00 and 18:57:00--18:58:00\,UT) and the low-B source (bottom row, 18:55:00--18:57:00\,UT). 
The red contour shows the 20\% ROI level from Hinode/XRT at 18:56:25\,UT, and the violet contour indicates the 10\% brightness level from the 5.79\,GHz EOVSA  image at 18:55:44\,UT.
 \label{Fig:Alfven velocity}
}
\end{figure*}

Figure\,\ref{Fig:Alfven velocity} presents the spatial distributions of the Alfvén speed (\( v_{\mathrm{A}} \)), plasma beta (\( \beta \)), and the plasma-to-gyrofrequency ratio (\(\alpha = f_{\mathrm{pe}} / f_{\mathrm{Be}} \)) for two intervals: 18:55:00–18:57:00 UT (top and bottom rows) and 18:57:00–18:58:00 UT (middle row; the bottom row shows the results obtained for the low-$B$ source; see Sect.\,\ref{S_reg_parm_map} and the bottom row of Fig.\,\ref{Fig:B_n_med}).  Within the ROI (red contour), plasma beta remains well below unity in the main source, indicating a magnetically dominated environment. The Alfvén speed ranges there from \( 2 \times 10^8 \) to over \( 8 \times 10^8\,\text{cm\,s}^{-1} \). The ratio \(\alpha \) stays around 1 across most of the ROI. In contrast, in the secondary, "low-$B$" source, the Alfvén velocity is about ten times smaller than in the main source, the plasma $\beta$ is about 1, and the frequency ratio is well above 1.  

Despite the general regularity seen among the parameter ranges between the two intervals, the internal structure of the flaring volume is highly inhomogeneous. Localized variations in \( v_{\mathrm{A}} \), \( \beta \), and \( f_{\mathrm{pe}} / f_{\mathrm{Be}} \) reveal complex plasma and magnetic configurations, which are crucial for revealing the effects of wave propagation, reconnection physics, and particle acceleration mechanisms.

\section{Discussion} \label{sec:discussion}

This study reports 3D reconstruction of the magnetic field and other physical parameters in the coronal source of the 2021-05-07 flare obtained using a combination of imaging microwave spectroscopy  and stereoscopic SXR data. To infer the magnetic field, we employed a gyrosynchrotron spectral model fitting of the EOVSA imaging spectroscopy data, which also provides thermal number density and nonthermal electron properties. 

The microwave data reveal a complexity of the flare source with at least two distinct microwave sources at many LOSs and time frames. At the beginning of the flare, roughly 18:55--18:58\,UT, the data allow for these sources to be disentangled with distinctly different microwave-inferred parameters. One of them displays the thermal number density matching that inferred from the SXR data, which permitted us to identify this microwave source with the SXR source. Therefore, the 3D coordinates derived from the SXR stereoscopic data can be applied to the magnetic field and other parameters inferred from the microwave spectral model fitting, providing us with previously unavailable evolving 3D distributions of the coronal magnetic field in the flaring region, along with a number of other inferred parameters (notably the thermal and nonthermal number densities). This highly important new information on the magnetic field in the flaring volume permitted us, in particular, to investigate dependence of this magnetic field on the height and put these data in the context of other measurements and models of the coronal magnetic field.

Moreover, having access to 3D data on both the magnetic field and plasma density enables the construction of 3D distributions of key parameters that govern plasma dynamics, such as the Alfvén velocity and plasma beta.
These quantities play a fundamental role in the physics of flares. 
The plasma beta quantifies the relative importance of thermal and magnetic pressure. 
 As emphasized by \citet{Gary2001SoPh}, the assumption \( \beta \ll 1 \) is not always valid throughout the corona and accurate magnetohydrodynamic modeling requires knowledge of \(\beta\), especially in regions where it approaches or exceeds unity.  
 Properly accounting for high-beta environments is essential, as they  could potentially invalidate the force-free approximation \citep{2012LRSP....9....5W}. Moreover, the plasma beta plays a pivotal role in both the theoretical understanding and numerical modeling of magnetic reconnection and current sheet formation \citep{Chen2020}.
The Alfvén speed sets the characteristic scales for MHD wave propagation and reconnection rates. In models of particle acceleration \citep[e.g.,][]{Arnold2021}, the Alfvén speed often plays a central role: it defines the normalization of fundamental physical scales (including energy, time, and length), while governing the rate of electron acceleration and influencing the spectral slope (i.e., the hardness) of the resulting electron energy distribution.
 
Our ability to derive these quantities in 3D using coordinated microwave and X-ray data offers a major leap in connecting observed structures with the plasma conditions responsible for energy release and particle acceleration. While many observational studies are limited to 2D projections or LOS integrals, spatially resolved 3D diagnostics allow us to identify localized variations essential for flare dynamics.

These 3D distributions provide important, earlier unavailable, observational constraints for theoretical models of flare energy release and particle acceleration \citep[e.g.,][]{Arnold2021} and in a broader context of the magnetohydrodynamical (MHD) models of the solar corona. For example, the Alfvén velocity is the key parameter in the theory of the MHD waves, which is foundational to coronal seismology \citep[e.g.,][]{McLaughlin2011SSRv..158..205M,2024RvMPP...8...19N}. Traditionally, the coronal seismology relied on simplified models, such as uniform cylindrical structures \citep{1982ZaitsevSvAL....8..132Z,Edwin1983SoPh...88..179E}, to interpret coronal oscillations. These models typically assume spatially uniform magnetic field and plasma density, enabling analytical solutions that (as compared to observations) can yield mean or weighted values of the Alfvén velocity and other relevant parameters in the analyzed loop \citep[e.g.,][]{2024RvMPP...8...19N}.

Meanwhile, our results show that both the Alfvén speed and plasma beta vary significantly within the flare region, as illustrated in Fig.~\ref{Fig:Alfven velocity}, which is in line with several earlier reports \citep{Verwichte2013ApJ...767...16V,Cho2017ApJ...837L..11C}. These spatial variations reveal the nonuniformity of the magnetic field and plasma density  in the flaring loop. The observed complexity challenges the simplified seismological models and calls for a more sophisticated advanced modeling approaches. 
Incorporating these nonuniformities into coronal seismology is crucial for improving the realism of the MHD models, which is needed for reliable coronal diagnostics. 

Finally, we computed a 3D distribution of the plasma-frequency-to-gyrofrequency ratio, $\alpha$, which turned out to be equal to about 1 in our flaring region. This parameter controls the coherent emission mode, which can be generated in the plasma. Roughly speaking, if this ratio is smaller than 1, then the electron cyclotron maser (ECM) emission may dominate; meanwhile, in the opposite case, the plasma emission may dominate provided that some instability conditions are met \citep{Fl_Meln_1998,2024NatAs...8...50Y}. In our case, we have both $\alpha<1$ and $\alpha>1$; therefore, conditions for both ECM and plasma emission can be met.

To conclude, we have reported a new powerful methodology for recovering 3D magnetic field and its associated parameters such as the Alfvén speed and plasma beta. We applied this methodology to a solar flare jointly observed with a set of multiwavelength instruments. The derived 3D distributions of the parameters can now be used as important observational constraints in various aspects of the flare modeling.\\

\begin{acknowledgements} 
EOVSA was designed, built, and is now operated by the New Jersey Institute of Technology (NJIT) as a community facility. EOVSA operations are supported by NSF grant AGS-2436999 to NJIT.
TK was supported by DFG grant eBer-24-58553. SY is supported by NASA grant 80NSSC24K1242 and NSF grant AGS-2334931 to NJIT. GF was supported in part by NSF grant AGS-2425102, and NASA grant 80NSSC23K0090 to New Jersey Institute of Technology.
\end{acknowledgements}

\bibliography{sample_and_fleishman}
\bibliographystyle{aa}
\bibpunct{(}{)}{;}{a}{}{,} 
\end{document}